\documentclass[sigconf,screen,nonacm]{acmart}

\usepackage{enumitem}

\AtBeginDocument{%
  }

\setcopyright{rightsretained}
\acmConference[]{}{}{}
\acmYear{2025}\copyrightyear{2025}
\acmDOI{}
\acmISBN{}
\acmBooktitle{}

\usepackage{graphicx}
\usepackage{xcolor}
\usepackage{colortbl}
\usepackage{xspace}

\newcommand{\orion}{\textsc{Orion}\xspace}

\begin{document}

\title{Orion: Fuzzing Workflow Automation}

\author{Max Bazalii}
\authornote{Both authors contributed equally to this research.}
\affiliation{%
  \institution{NVIDIA}
  \city{Santa Clara}
  \country{USA}
}
\email{mbazalii@nvidia.com}

\author{Marius Fleischer}
\authornotemark[1]
\affiliation{%
  \institution{NVIDIA}
  \city{Santa Clara}
  \country{USA}
}
\email{mfleischer@nvidia.com}

\begin{abstract}
Fuzz testing is one of the most effective techniques for finding software vulnerabilities. 
While modern fuzzers can generate inputs and monitor executions automatically, the overall workflow, from analyzing a codebase, to configuring harnesses, to triaging results, still requires substantial manual effort.
Prior attempts focused on single stages such as harness synthesis or input minimization, leaving researchers to manually connect the pieces into a complete fuzzing campaign.

We introduce \orion, a framework that automates the the manual bottlenecks of fuzzing by integrating LLM reasoning with traditional tools, allowing campaigns to scale to settings where human effort alone was impractical.
\orion uses LLMs for code reasoning and semantic guidance, while relying on deterministic tools for verification, iterative refinement, and tasks that require precision.
Across our benchmark suite, \orion reduces human effort by 46–204$\times$ depending on the workflow stage, and we demonstrate its effectiveness through the discovery of two previously unknown vulnerabilities in the widely used open-source \texttt{clib} library.
\end{abstract}



\maketitle

\section{Introduction}

{\emergencystretch 3em
Fuzzing is one of the most effective techniques for vulnerability discovery. 
By executing programs with large numbers of random, malformed, or unexpected inputs, fuzzing has exposed tens of thousands of real-world bugs. 
Google's OSS-Fuzz~\cite{ossfuzz} service alone has reported more than 50,000 vulnerabilities across over 1,000 open-source projects, while syzkaller~\cite{syzbot} has uncovered more than 8,000 kernel bugs. 
These results highlight fuzzing's importance both in research and in production environments.
}

However, scaling fuzzing to large production codebases remains a major challenge.
Modern fuzzers automate input generation and coverage feedback once a campaign is running, but the broader workflow surrounding a fuzzing campaign is still heavily manual. 

Analysts must first examine the codebase to identify promising fuzz targets. 
They then implement \textit{fuzz harnesses}, small programs that accept fuzzer inputs and invoke target functions with the correct arguments. 
Additionally, harness construction is often coupled with \textit{seed generation}, the preparation of example inputs that illustrate valid input structures and allow the fuzzer to exercise deeper execution paths.

Post-execution, the analyst must analyze results: triaging crashing inputs, interpreting stack traces, and identifying root causes.
Finally, the analyst must write patches for all discovered bugs.
These crash analysis and patch generation tasks are particularly time-consuming, requiring a deep understanding of the target program and codebase.

This dependency on manual effort creates a major bottleneck for scaling fuzzing to large production codebases, representing the primary focus of this work.

Prior research has explored automation for individual workflow components, but each approach comes with limitations. 
Traditional automated patching systems like GenProg~\cite{genprog} and SemFix~\cite{semfix} face fundamental scaling concerns with large codebases.
Existing fuzzing automation tools impose restrictive dependencies:  Utopia~\cite{utopia} requires unit test availability, Winnie~\cite{winnie} depends on existing exemplar programs for harness generation, and Skyfire~\cite{skyfire} relies on pre-existing seed corpora.
Critically, these approaches focus exclusively on single workflow steps rather than end-to-end processes. Substantial manual effort is still required to connect these pieces into an end-to-end fuzzing campaign.


Recent advances in large language models (LLMs) suggest a new direction. 
LLMs demonstrate strong capabilities in code understanding, reasoning, and generation. 
When combined with agentic frameworks, they can perform long-horizon planning, decompose multi-step tasks, and iteratively refine their own outputs. 
This makes them natural candidates for automating complex engineering workflows such as fuzzing. 
Several works have begun exploring this direction—for example, harness generation~\cite{googleHarnessGeneration, promptFuzzingFuzzDriver, ckgFuzzer}, seed creation~\cite{seedmind, psgSeedGen}, or crash analysis~\cite{coderovers}—but all focus on isolated stages. 
No existing system provides an integrated, end-to-end solution.

We present \orion, a framework that automates the fuzzing workflow from start to finish by combining LLM agents with deterministic analysis tools. 
The core design principle is to leverage each component for what it does best: LLMs provide semantic reasoning about codebases and assist in creative tasks such as harness design or seed inference, while traditional tools supply reliable checks through compilation, execution, and static analysis. 
Outputs from LLM agents are validated wherever possible, and errors are fed back into iterative refinement loops. 
This mitigates the unreliability of probabilistic models while exploiting their strengths in code reasoning and generation. 

\orion mirrors the workflow of human fuzzing experts. 
It begins by transforming the target codebase into a structured knowledge base, enabling precise retrieval of relevant context for subsequent agents. 
Specialized subsystems then carry out each stage of the workflow: target identification, seed generation, harness creation, crash triage, and patch suggestion. 
Agents are equipped with the ability to invoke external tools, run code, and collect feedback, and employ prompting strategies such as self-consistency and self-reflection to improve reliability. 
Through this design, \orion replaces the manual bottlenecks of fuzzing with automated, tool-supported agents, scaling fuzzing to settings where human effort previously made it impractical.

We evaluate \orion using a custom benchmark across two open-source and one proprietary library. Across these targets, \orion reduces required human effort by 46–204$\times$ depending on the workflow stage. 
During evaluation, it also discovered two previously unknown vulnerabilities in the widely used \texttt{clib} library~\cite{clib}, both responsibly disclosed to developers. 
These results demonstrate that combining LLM agents with deterministic verification can make end-to-end fuzzing automation practical.

In summary, we make the following contributions:

{\emergencystretch 3em
\begin{enumerate}
    \item We introduce \orion, the first end-to-end fuzzing workflow automation framework that integrates LLM agents with traditional tools.
    \item We design techniques to improve LLM reliability in this setting, including tool-assisted verification, feedback-driven refinement, and prompting strategies such as self-consistency and self-reflection.
    \item We implement and evaluate \orion on real-world codebases, showing substantial reductions in required human effort and the discovery of two zero-day vulnerabilities.
\end{enumerate}
}



\section{Background}

\subsection{Fuzzing 101}
\label{sec:fuzzing_101}

Fuzzing is a widely used dynamic analysis technique for vulnerability discovery. 
It executes target programs with automatically generated inputs while monitoring for abnormal behavior or crashes. 
When a crash occurs, the fuzzer records the crashing input and stack trace for later analysis.

Modern fuzzers employ coverage guidance to increase exploration efficiency~\cite{aflplusplus}. 
Inputs that trigger new execution paths are retained in the fuzzer's corpus and subsequently mutated, allowing the fuzzer to incrementally learn the structure of valid inputs rather than starting from scratch. 
To detect bugs beyond crashes, fuzzers rely on runtime instrumentation, or \textit{sanitizers}, that identify invalid program states (e.g., out-of-bounds accesses, use-after-free) and deliberately terminate execution with diagnostic messages. 
Sanitizers extend fuzzing's reach to memory safety, thread safety, undefined behavior, and memory leak detection.

Effective fuzzing requires careful interface selection. 
Fuzzing every interface in a large codebase is infeasible, and many interfaces are irrelevant from a security perspective. 
Analysts typically focus on attacker-accessible interfaces or those with high security relevance (e.g., authentication logic), guided by existing threat models. 
Since fuzzing identifies issues through crashes, the chosen targets must include code where invalid states can manifest as observable failures (e.g., invalid pointer dereferences).

Harnesses are required to connect fuzzers to target programs. 
A fuzz harness receives byte buffer inputs from the fuzzer, transforms them into the program's expected format, and invokes the target interface. 
Harness quality has a direct impact on effectiveness: poor harnesses can cause spurious crashes, misuse interfaces, or exhaust resources, thereby obscuring real vulnerabilities. 
High-quality harnesses perform proper initialization and dependency setup, execute the target interface correctly, and include teardown logic to restore program state. 
This ensures reproducibility across iterations and minimizes runtime overhead, enabling fuzzers to operate at maximum speed.

Seed corpora further improve fuzzing effectiveness. 
While fuzzers can start with an empty corpus, providing example inputs accelerates progress by illustrating valid input structures~\cite{sokFuzzingEvals}. 
High-quality seed corpora contain diverse inputs that exercise a wide range of program behaviors and edge cases, enabling the fuzzer to focus on deeper exploration rather than format inference. 

Fuzzers nevertheless struggle with input constraints that are difficult to satisfy through random mutation, such as checksums or constant value comparisons. 
Although fuzzers favor common boundary values (e.g., -1 for integers), complex input validation logic often requires crafted seeds or harness logic that encodes expected values. 
Such guidance is critical for enabling the fuzzer to reach code behind hard-to-satisfy conditions.

At the end of a campaign, fuzzers report coverage statistics and discovered crashes, each accompanied by the triggering input and stack trace. 
However, the crash site may differ from the actual defect location, since corrupted state can propagate before triggering a visible failure. 
A single defect can produce multiple crashes (e.g., a missing NULL assignment after freeing memory may appear as both a use-after-free and a double-free). 
Consequently, crash triaging and root cause analysis remain essential steps before fixing bugs.





\subsection{Human Fuzzing Workflow}
\label{sec:human_fuzzing_workflow}
Human analysts typically follow a structured workflow when fuzz\-ing target programs. 

The process begins with \textbf{target selection}. 
Analysts leverage threat models and source code analysis to identify interfaces that balance high bug likelihood with attacker accessibility. 
These interfaces are prioritized for fuzzing campaigns.

Next, analysts construct \textbf{fuzz harnesses} for selected targets. 
This requires detailed knowledge of input formats and interface dependencies. 
Harnesses must provide appropriate initialization, interface invocation, and cleanup logic. 
Lightweight and reliable harnesses are critical for campaign success.

Before execution, analysts design \textbf{seed corpora} considering target interface behavior and expected harness input formats.
The goal is to create diverse corpora that demonstrate expected input formats and edge cases.

Analysts then execute \textbf{fuzzing campaigns}, typically the only phase that runs autonomously once configured. 
Fuzzers operate until they uncover crashing inputs, which are logged with execution details. 

When crashes occur, analysts conduct \textbf{bug triage and patching}. 
Triaging involves identifying root causes, correlating multiple manifestations of the same bug, and assessing severity. 
This requires interpreting stack traces, analyzing source code, and replaying crashing inputs with debugging tools. 
Once the underlying issue is identified, analysts develop and test patches, often requiring several iterations before resolution.

This workflow illustrates that only the execution phase is substantially automated today; the surrounding tasks still depend heavily on human expertise.



\subsection{LLMs and LLM Agents}
\label{sec:llms}

Large language models (LLMs) are transformer-based neural networks trained on massive text corpora to predict token sequences, thereby capturing statistical patterns of natural language and source code~\cite{transformer, retrievalAugmentedGenerationSurvey}. 
After pretraining, instruction tuning and alignment (e.g., RLHF) enable them to follow prompts for tasks such as question answering, code generation, and multi-step reasoning. 
Benchmarks such as SWE-bench~\cite{sweBench} and HumanEval~\cite{humaneval} demonstrate LLMs' ability to understand program semantics and generate functional code. 
These capabilities have enabled widespread adoption of LLMs in development workflows (e.g., IDE assistants~\cite{cursor, windsurf}) and in agent-based frameworks for software engineering~\cite{codex, devin, claudeCode}. 
Applications to security-specific tasks are discussed further in Section~\ref{sec:related_work}.

However, LLMs exhibit significant limitations requiring careful system design.
Knowledge retrieval errors frequently occur, particularly for fine details crucial in code understanding, often leading to hallucinations where LLMs generate incorrect information~\cite{retrievalAugmentedGenerationSurvey}.
Context window constraints restrict the amount of text they can process, and large contexts often suffer from the \textit{needle-in-the-haystack} problem, where critical details are overlooked~\cite{needleInTheHaystack}. 
LLMs also struggle with counting and arithmetic, and their probabilistic nature produces inconsistent outputs even for identical inputs. 
These weaknesses complicate reliable use in security workflows, despite mitigation techniques such as self-consistency~\cite{selfConsistency}.

LLM agents extend base models with tool usage, external memory, and planning capabilities~\cite{wengAgent}. 
In this paradigm, the LLM serves as the agent's reasoning core while tools enable concrete actions beyond text generation (e.g., invoking command-line utilities). 
Such approaches have been applied to software development~\cite{claudeCode, codex, devin} and security analysis~\cite{xbow}. 
However, agents amplify underlying LLM weaknesses: tool misuse can corrupt environments (e.g., accidental file deletion), and iterative reasoning can propagate early mistakes across multiple steps. 
These error cascades make it difficult to rely on LLM agents for tasks that demand precision, such as crash triage or automated patching in security testing.

\section{\orion Design}



\begin{figure*}[t]
    \centering
    \includegraphics[width=\textwidth]{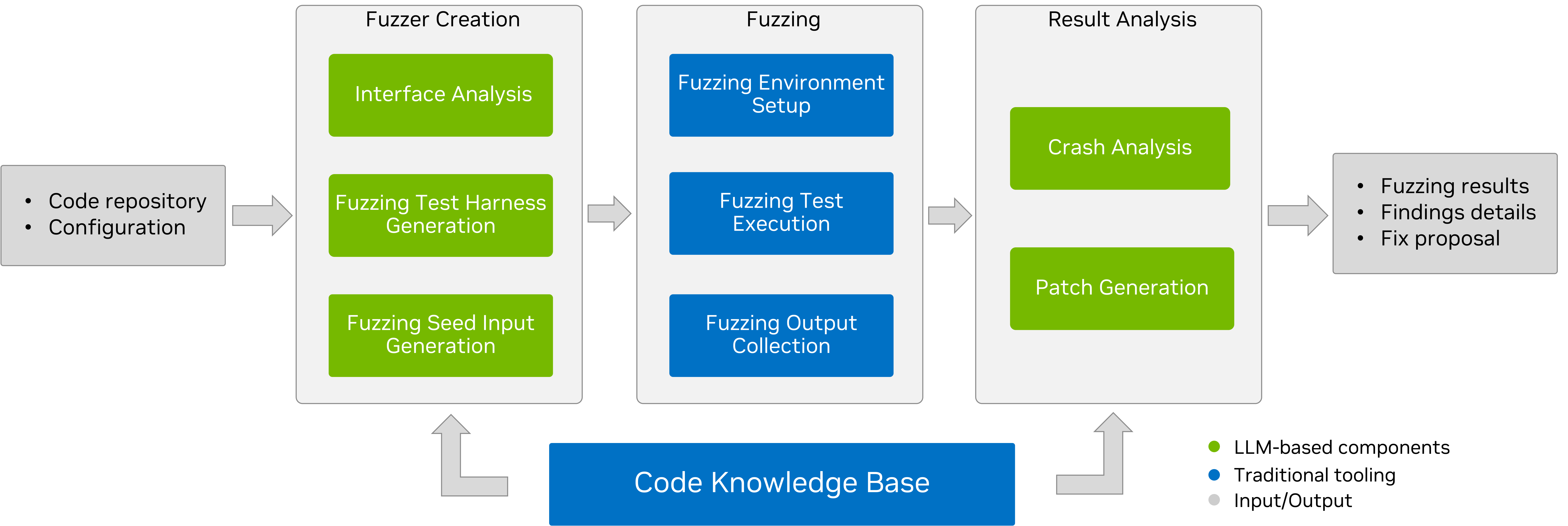}
    \caption{Overview of the \orion fuzzing workflow automation pipeline. }
    \label{fig:orion_workflow}
\end{figure*}

We present \orion, a framework for automating the fuzzing workflow by combining LLM-based agents with deterministic tools. 
The design mirrors the workflow of a human analyst (Figure~\ref{fig:orion_workflow}) and enables fuzzing campaigns to scale with minimal manual effort. 
\orion supports pre-fuzzing tasks including target identification (Section~\ref{sec:target_identification}), seed generation (Section~\ref{sec:seed_generation}), and harness construction (Section~\ref{sec:harness_generation}), as well as post-fuzzing tasks such as crash triage (Section~\ref{sec:triaging_crash_analysis}) and patch generation (Section~\ref{sec:patching_agent}). 
Each stage is handled by a dedicated agent, which decomposes the workflow into simpler subtasks and allows independent optimization of components. 
A codebase indexer (Section~\ref{sec:code_base_indexing}) underpins all stages by enabling targeted information retrieval.

\subsection{Overview of the \orion Workflow}
\label{sec:orion_workflow}

\orion takes as input the target project's source code and build configuration (e.g., compilation commands, compiler flags, agent parameters). 
It begins by constructing a codebase index (Section~\ref{sec:code_base_indexing}) to support targeted queries. 
The target identification stage (Section~\ref{sec:target_identification}) then selects candidate interfaces for fuzzing by ranking non-static functions likely to contain fuzzer-triggerable bugs.

For each selected function, \orion generates seed inputs (Section~\ref{sec:seed_generation}). 
Seeds are crafted to address input constructs that fuzzers typically struggle with (e.g., checksums) and to provide structural specifications used during harness generation (Section~\ref{sec:harness_generation}).

Harness generation (Section~\ref{sec:harness_generation}) involves two agents: a dependency analysis agent (Section~\ref{sec:harness_generation_dependency_analysis}) and a harness synthesis agent (Section~\ref{sec:harness_generation_harness_generation}). 
The dependency agent identifies required setup and teardown procedures, as well as header file dependencies, while the harness agent constructs compilable fuzz drivers compatible with generated seeds. 
Compiler feedback ensures harness validity and enables iterative refinement when errors are detected.

The resulting harnesses and seeds are passed to the fuzzing execution infrastructure (Section~\ref{sec:fuzzing_execution}), which runs the fuzzer, monitors coverage, and records crashes. 
Coverage feedback can trigger harness or seed adjustments, while crashes are analyzed by downstream agents.

When crashes occur, the triage agent (Section~\ref{sec:triaging_crash_analysis}) filters out harness-related issues, determines root causes, and produces minimal reproducers. 
These reproducers support bug understanding and serve as validation inputs for the patching agent (Section~\ref{sec:patching_agent}), which proposes candidate patches by analyzing the faulty code. 
Proposed patches are validated through compilation and replay against reproducers before producing a final patch diff for application.

\noindent
\textbf{Challenges. }
The design of \orion highlights several challenges that must be addressed to achieve reliable automation.

First, as discussed in Section~\ref{sec:llms}, LLMs possess broad world knowledge but cannot retrieve it reliably at fine granularity. 
This is problematic when reasoning about source code, where small details are often critical. 
Moreover, proprietary codebases and documentation are not part of LLM pretraining data, so \orion cannot rely on pretrained knowledge alone and must supply relevant context. 
The key challenge is determining which information to provide: excessive context can confuse the model, while insufficient context can lead to fabricated outputs. 
\orion must therefore balance between overwhelming the model and omitting important details.

Second, LLM agents operate with broad action spaces, which increases flexibility but also risk. 
Incorrect reasoning traces may yield invalid outputs or corrupt the runtime environment~\cite{replitIncident}. 
While agents offer benefits through semantic reasoning beyond what deterministic tools can capture, it is critical to assign them tasks that match their strengths and to enforce oversight mechanisms that limit harmful behavior.

Finally, LLMs are inherently stochastic (Section~\ref{sec:llms}), producing different outputs even for identical inputs. 
Without strong oracles to verify correctness, these random variations can propagate through the workflow, invalidating subsequent analyses and undermining reproducibility. 
This motivates the design mitigations described below.

\noindent
\textbf{Techniques. } 
We now describe how \orion addresses the above challenges.

To overcome context retrieval limitations, \orion employs a codebase indexer built on compiler tooling. 
The indexer supports fine-grained queries for specific program elements (e.g., function or type definitions, header declarations), enabling the system to supply the LLM with precisely the context required for each task. 
This reduces both information overload and the risk of missing crucial details.

To balance LLM strengths with reliability, \orion integrates them with deterministic tools. 
Tasks that can be handled precisely by existing tools (e.g., code metric calculation) are delegated entirely to those tools. 
For tasks requiring semantic reasoning, LLMs operate in conjunction with traditional tools that both provide structured context and validate outputs. 
Tool feedback loops prevent erroneous results from propagating further in the workflow.

Finally, \orion improves reproducibility through consensus-\linebreak based prompting techniques. 
Self-consistency~\cite{selfConsistency} issues multiple queries with identical input and aggregates results, selecting the majority outcome to mitigate stochastic variation. 
Self-reflection~\cite{selfReflection} allows the LLM to critique its own reasoning process and revise outputs based on detected errors. 
These mechanisms, applied across all workflow stages, enhance reliability and reduce the impact of individual model errors.

\subsection{Code Base Indexing}
\label{sec:code_base_indexing}



The codebase indexer is the first component executed by \orion and underpins all subsequent workflow stages by addressing two key challenges in context retrieval for LLMs.

First, LLMs have limited context windows, which are too small to accommodate most production codebases (Section~\ref{sec:llms}). 
A common mitigation is to split input into smaller chunks and select those most relevant to the current query. 
For code analysis, however, chunking must respect semantic boundaries to preserve meaning. 
\orion therefore uses function boundaries as natural chunking units, since functions are self-contained logical entities and common fuzzing targets are function entry points. 
This design enables precise retrieval and supports reasoning at the granularity of individual interfaces.

Second, because proprietary codebases and project-specific libraries are not part of LLM pretraining data (Section~\ref{sec:orion_workflow}), \orion must supply all necessary context explicitly. 
This includes function signatures, definitions, type declarations, header file contents, and caller–callee relationships. 
Accurate retrieval of this information is critical, as irrelevant or missing context directly reduces the reliability of LLM reasoning.

The indexer addresses these requirements by parsing source code with the compiler toolchain. 
It extracts function metadata (declarations, definitions, signatures, scope), constructs a global call graph, and builds a type index associated with header files. 
These artifacts enable targeted queries such as retrieving functions defined in specific headers (target identification, Section~\ref{sec:target_identification}), resolving types by header (harness generation, Section~\ref{sec:harness_generation_harness_generation}), or recursively enumerating callees (dependency analysis, Section~\ref{sec:harness_generation_dependency_analysis}). 
By grounding LLM prompts in compiler-derived information, the indexer ensures that \orion can provide the model with precise and relevant context without exceeding its limits.

\subsection{Target Identification}
\label{sec:target_identification}

As described in Section~\ref{sec:human_fuzzing_workflow}, human analysts begin fuzzing by selecting interfaces that balance attacker relevance and expected bug density. 
This typically requires external threat models, which limits applicability to well-studied systems. 
To support broader deployment, \orion instead focuses on identifying interfaces that maximize fuzzer effectiveness, ensuring practical results even when threat models are unavailable.

Because fuzzers primarily expose sanitizer-detectable bugs, the goal is to prioritize interfaces with high likelihood of containing such vulnerabilities. 
\orion estimates this likelihood using a set of independent metrics (Section~\ref{sec:target_identification_metrics}) that evaluate functions individually. 
Metric outputs are then combined into a composite ranking from which \orion selects the top candidates within the available fuzzing budget.

The metric set captures diverse code properties correlated with vulnerability risk, though it is not exhaustive and no single metric can definitively identify vulnerable interfaces. 
\orion{}'s design allows new metrics to be introduced or existing subsets to be selected, enabling customization for different codebases.

Several metrics rely on LLM analysis, raising concerns about reliability. 
To address this, \orion applies self-consistency and self-reflection (Section~\ref{sec:llms}): multiple evaluations are aggregated to mitigate stochastic variation, and outputs are iteratively reviewed and refined to reduce errors.

Finally, since metrics produce heterogeneous outputs (e.g., numerical scores, binary indicators, text patterns), direct aggregation is infeasible. 
\orion therefore prompts the LLM to synthesize results into a ranked list of fuzzing targets.

\subsubsection{Target Identification Metrics}
\label{sec:target_identification_metrics}
\orion applies a set of metrics to prioritize candidate fuzzing targets. Each captures a distinct property correlated with vulnerability risk.

\noindent
$\blacktriangleright$ \textit{Cyclomatic Complexity} measures the number of linearly independent paths through a function. 
Higher complexity correlates with increased likelihood of defects. 
This metric is computed directly using the codebase indexer and static analysis tooling.

\noindent
$\blacktriangleright$ \textit{Internal Function Calls} counts how frequently a function is invoked by others. 
Bugs in frequently used functions can have greater security impact than those in rarely used code. 
This metric is derived from the call graph.

\noindent
$\blacktriangleright$ \textit{Lines of Code (LOC)} serves as a simple proxy for complexity: longer functions are harder to reason about and more likely to contain errors. 
LOC is measured from function bodies retrieved by the indexer.

\noindent
$\blacktriangleright$ \textit{Callgraph Size} measures the number of functions reachable from the analyzed function. 
Large reachable sets indicate higher structural complexity and broader potential impact of vulnerabilities. 
This is computed from the global call graph.

\noindent
$\blacktriangleright$ \textit{Dangerous Expressions} identifies low-level constructs such as pointer arithmetic, manual memory management, and bit-level operations, which are historically associated with memory safety issues. 
These are detected using LLMs, which can leverage surrounding context to distinguish benign from potentially dangerous uses.

\noindent
$\blacktriangleright$ \textit{Sink Functions} captures functions commonly linked to security vulnerabilities (e.g., \texttt{memcpy}, \texttt{strcpy}, \texttt{malloc}). 
\orion builds on Fuzz-Introspector's CVE-based function set~\cite{fuzzIntrospector, cweData}, extending it with LLM support to detect wrapper functions or project-specific variants, reducing manual configuration effort.

\noindent
$\blacktriangleright$ \textit{Parsing Functions} identifies functions that parse structured inputs, which are prone to errors due to format complexity, recursion, and edge cases. 
LLMs are used here as well, since names alone are insufficient to recognize project-specific parsers or unconventional naming schemes.



\subsection{Seed Generation}
\label{sec:seed_generation}

The seed generation phase receives target interfaces from the previous step and produces high-quality seeds that improve initial coverage, demonstrate valid input formats, and address constraints that fuzzers typically struggle to satisfy (e.g., input validation, checksums; Section~\ref{sec:fuzzing_101}). 
Unlike human analysts, who create seeds after writing harnesses, \orion generates seeds first. 
Both tasks require understanding the function's input space, but harness construction additionally involves dependency resolution and correct invocation. 
By generating seeds first, \orion reduces harness complexity by providing parsing specifications and concrete examples, which improves LLM reliability in both stages (Section~\ref{sec:llms}).

The seed generation agent analyzes each target function to characterize its inputs and behavior, then crafts diverse seed inputs covering expected paths, edge cases, and error conditions. 
Seeds are produced as scripts that generate input files rather than as raw binary data, since LLMs are better at code generation than binary output. 
Execution of these scripts provides natural feedback: errors are detected and used to refine outputs through iterative self-reflection. 
The final outputs consist of (1) concrete seed files, (2) textual descriptions of seed formats, and (3) code analysis results.

Preliminary code analysis is performed before seed construction to capture relevant input and behavioral properties. 
This combines information from the codebase indexer with reasoning from LLM agents. 
Each analysis produces written findings that include semantic details (e.g., from comments or developer conventions) often missed by static tools. 

\noindent
$\blacktriangleright$ \textit{Function Signature Analysis} extracts function prototypes and parameter details, including expected value ranges, implicit relations, ownership semantics, and default values.  

\noindent
$\blacktriangleright$ \textit{Input Surface Analysis} examines input channels beyond function arguments (e.g., files, globals, network, IPC) and identifies constraints for each.  

\noindent
$\blacktriangleright$ \textit{Control Flow Analysis} summarizes behavior by identifying branch\-es, loops, and recursive calls, distinguishing normal paths from error-handling paths.  

\noindent
$\blacktriangleright$ \textit{Memory Management Analysis} identifies heap/stack interactions, allocation sizes, and buffer operations, highlighting potential risks such as out-of-bounds accesses or use-after-free.  

\noindent
$\blacktriangleright$ \textit{Error Handling Analysis} inspects how the function detects errors, propagates faulty state, and performs recovery or graceful failure.  

\noindent
$\blacktriangleright$ \textit{Call Dependency Analysis} explores direct, indirect, and function-pointer-based calls to understand dependencies beyond static call graphs.  

\noindent
$\blacktriangleright$ \textit{Coverage Goal Analysis} highlights code regions most important to exercise (e.g., critical conditionals, memory operations, complex branches) and produces a prioritized coverage list.  

\noindent
$\blacktriangleright$ \textit{Vulnerability Analysis} provides heuristic estimates of potential fuzzer-detectable bugs (e.g., integer overflows, double frees, TOCTTOU issues). 
The goal is not precise detection~\cite{codeLLMVulnDetection}, but to provide starting points that guide seed generation toward high-value paths.


\subsection{Harness Generation}
\label{sec:harness_generation}

Given the selected target interfaces (Section~\ref{sec:target_identification}) and seeds with analysis results (Section~\ref{sec:seed_generation}), the harness generation phase constructs fuzzing harnesses compatible with both the seeds and the chosen fuzzing engine. 
High-quality harnesses must (i) correctly initialize the environment required by the interface (e.g., global state, files, sockets), (ii) parse fuzzer input and populate corresponding data structures, (iii) perform teardown operations after invocation to prevent leaks and ensure repeatability, and (iv) minimize runtime overhead to maximize the number of fuzzing iterations achievable within a time budget.

\orion employs two agents to meet these requirements. 
The \textit{dependency analysis agent} examines how interfaces interact with their environment (other functions, filesystem, global variables) and outputs setup and teardown requirements for each interface. 
The \textit{harness generation agent} then produces harnesses for individual interfaces, guided by the dependency information and seed analysis. 
Compiler feedback loops ensure harnesses compile and execute correctly, while self-reflection is used to refine them into compact implementations that reduce per-iteration cost.

\subsubsection{Dependency Analysis Agent}
\label{sec:harness_generation_dependency_analysis}

The dependency analysis agent identifies relationships between interfaces by examining their interactions with the runtime environment, including accesses to global variables, files, environment variables, devices, and pointer-based inputs. 
Dependencies occur when one interface requires specific state that another creates (e.g., a file must exist before it can be opened). 
Teardown relationships occur when interfaces revert state changes (e.g., freeing memory, resetting variables). 
Understanding these relations requires reasoning about program semantics that traditional static analysis tools cannot fully capture; thus the dependency agent relies primarily on LLM reasoning, supported by compiler-derived code retrieval.

The agent operates in three passes:
\begin{enumerate}[leftmargin=*]
  \item \textbf{State extraction:} For each interface, collect expected environment state (read accesses), modifications (writes, allocations), and pointer-based parameter expectations.
  \item \textbf{Dependency matching:} Compare expectations against modifications across interfaces. Matches are recorded as dependencies, each accompanied by a brief textual rationale.
  \item \textbf{Teardown identification:} Identify interfaces that revert prior modifications (e.g., deallocation of memory, removal of files) and record them as teardown relationships with supporting explanations.
\end{enumerate}

The output is a structured list of setup and teardown requirements. 
Each requirement includes a description of the state to be established or reverted and the set of interfaces capable of fulfilling it.

\subsubsection{Harness Generation Agent}
\label{sec:harness_generation_harness_generation}

\begin{figure}[t]
    \centering
    \includegraphics[width=\columnwidth]{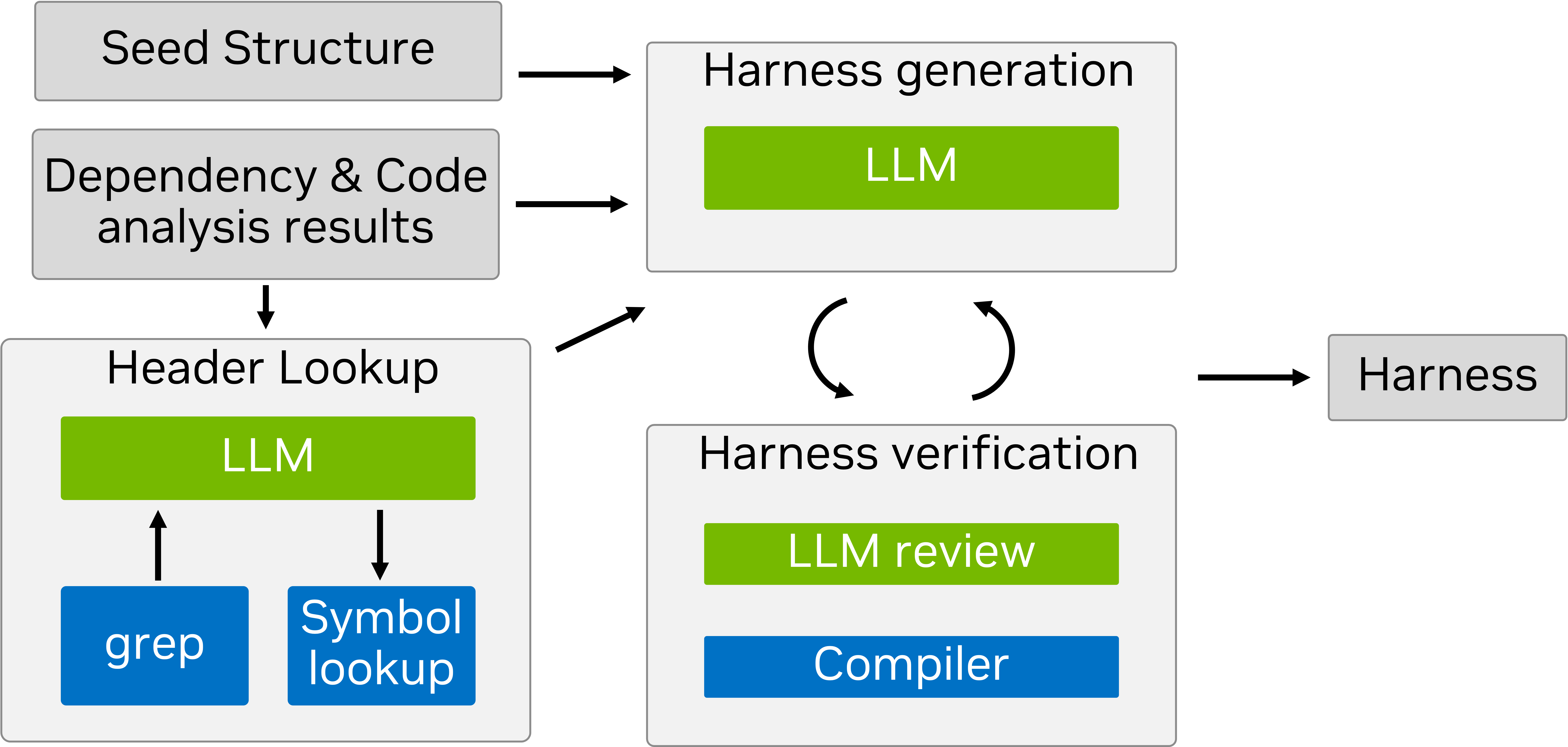}
    \caption{Workflow of the harness generation agent. }
    \label{fig:harness_agent}
\end{figure}

The harness generation agent produces compilable harnesses using the dependency analysis outputs, seed files, and seed analysis results (Figure~\ref{fig:harness_agent}). 
\orion targets userspace fuzzing and adopts the \texttt{LLVMFuzzerTestOneInput} interface, compatible with both libFuzzer and AFL++.

The agent begins by retrieving relevant header files through the codebase indexer and lightweight text search, as harness compilation requires correct type and interface declarations. 
It then assembles a structured prompt containing setup and teardown requirements, header files, available types and interfaces, seed structures, and input surface analyses. 
From this context, the LLM generates an initial harness candidate.

Harnesses are refined through iterative loops: self-reflection is used to review and adjust the generated code, followed by compiler checks to validate correctness. 
Compilation failures trigger additional self-reflection to ensure corrections do not reintroduce earlier errors. 
The process repeats until a valid harness is produced. 

The final output is a compilable harness aligned with the seed corpus and compatible with the selected fuzzing engine.



\subsection{Fuzzing Execution}
\label{sec:fuzzing_execution}

\begin{figure}[t]
    \centering
    \includegraphics[width=\columnwidth]{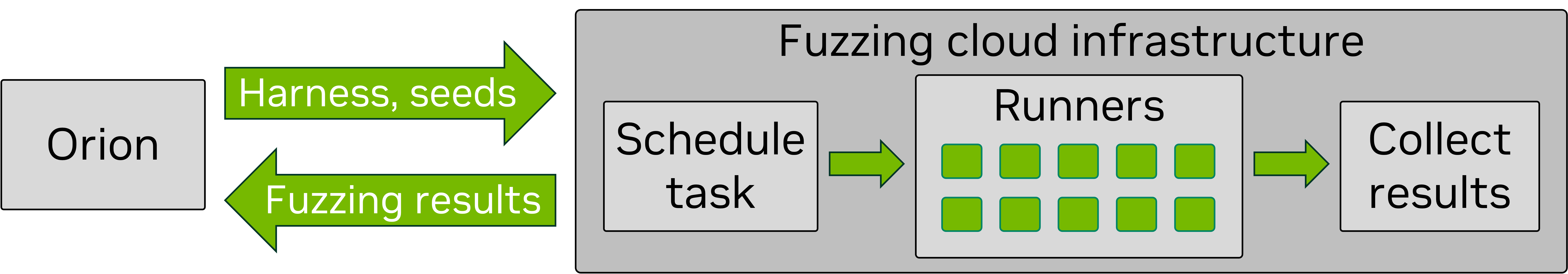}
    \caption{FuzzerHub execution model. }
    \label{fig:fuzzerhub}
\end{figure}

With harnesses and seed corpora prepared, the next step is to execute fuzzing campaigns against the selected interfaces. 
Although fuzzing engines run autonomously once launched, they require appropriate host setup and, in some cases, specialized hardware. 
Because infrastructure design is outside \orion{}'s core contributions, this section focuses on the execution platform capabilities needed to support the framework.

The fuzzing platform must:  
\begin{itemize}[leftmargin=*]
  \item accept automated submissions of harnesses and seed corpora;  
  \item execute fuzzing runs in isolated environments to avoid interference and ensure reproducibility;  
  \item collect results including coverage information, crash reports, and crashing inputs; and  
  \item return these results to \orion for downstream processing (Figure~\ref{fig:fuzzerhub}).  
\end{itemize}

Beyond executing campaigns, the collected statistics enable feedback loops: coverage data is compared against expectations to refine harnesses and seeds, while unexpected harness crashes can be diagnosed and corrected. 
Crashes attributable to the target interface, rather than the harness, transition \orion to the crash analysis and patching stages discussed next.


\subsection{Crash Analysis}

Scaling fuzzing campaigns produces large volumes of crashes, but these results are only useful if they can be triaged and acted upon. 
Raw crash reports often overwhelm developers, as many crashes are duplicates, false positives from harness issues, or lack sufficient detail for debugging. 
Without effective triage, the value of fuzzing campaigns is limited.

To address this, \orion integrates an automated crash analysis stage. 
The crash analysis agent, acting in place of a human analyst, filters out harness-related faults, clusters crashes that stem from the same underlying bug, and produces detailed reports. 
Each report includes minimal reproducers, stack traces, and contextual explanations of the suspected root cause. 
These outputs reduce the effort required by developers and provide a clean handoff to the patching stage, ensuring that fuzzing results can be remediated efficiently.

\subsubsection{Crash Triaging Agent}
\label{sec:triaging_crash_analysis}

The crash triaging agent automates two key objectives after crash discovery: (i) distinguishing harness-induced faults (false positives) from genuine interface crashes, and (ii) identifying likely root causes with minimal reproducers for the latter.

The agent receives crashing inputs, stack traces, and access to codebase exploration tools derived from the indexer. 
It analyzes these artifacts to generate minimal reproducers that exercise the same failure. 
Executing the reproducers against the target interface provides a direct feedback loop, confirming both root cause identification and the validity of the reproducer.

Reproducers serve a dual purpose: they give developers concise artifacts to validate and understand vulnerabilities, and they provide inputs for the patching agent. 
The triage stage outputs a structured report containing the suspected root cause, supporting stack trace information, and validated minimal reproducers.


\subsubsection{Patching Agent}
\label{sec:patching_agent}

The patching agent is the final stage of \orion{}'s workflow. 
Its goal is to propose candidate patches for vulnerabilities identified during crash triage. 
Effective patches should (i) eliminate the root cause without altering intended interface behavior, (ii) avoid introducing new vulnerabilities or regressions, and (iii) conform to project coding style and apply cleanly.

To approach these goals, the patching agent is equipped with tools for direct file inspection and modification, symbol lookup via the codebase indexer, compiler invocation, and reproducer execution. 
These capabilities allow the agent to interleave code exploration, code modification, and validation. 
Candidate patches are compiled and tested against the minimal reproducers; only those that build successfully and prevent the original crash are returned. 

The output of this stage is a patch proposal that passes automated validation. 
Developers may then review and apply the patch, modify it as needed, or reject it, ensuring that final acceptance remains under human control.




\section{Evaluation}


We evaluate \orion with respect to the following research questions:

\textbf{RQ1:} How effective is \orion at ranking interfaces using the identification metrics? (Section~\ref{sec:rq1})

\textbf{RQ2:} What is the quality of the fuzzing harnesses generated by \orion? (Section~\ref{sec:rq2})

\textbf{RQ3:} How much time does \orion save compared to human analysts? (Section~\ref{sec:rq3})

\textbf{RQ4:} Can \orion be applied to discover previously unknown vulnerabilities? (Section~\ref{sec:rq4})

\subsection{Benchmark Setup}
\label{sec:benchmark}

To study RQ1 and RQ2, we developed a benchmark spanning both open-source and proprietary projects. 
The open-source set includes \texttt{clib}~\cite{clib} (commit \texttt{6d96e533}) and \texttt{H3}~\cite{h3} (commit \texttt{d5af2344}), while the proprietary set consists of an NVIDIA QNX GPIO driver. 
This diversity allows us to assess \orion across multiple domains, code sizes, and coding styles, and to identify both strengths and limitations.

All experiments use GPT-4.1 as \orion{}'s primary reasoning model. 
For evaluation and judging tasks, we employ multiple LLMs as independent reviewers: Llama 3.1 405B, Llama 3.3 70B, and Llama 3.3 Nemotron Super 49B v1.

\subsection{Benchmark Design}
\label{sec:benchmark}

The benchmark is designed to provide systematic evaluation across multiple metrics, each capturing a distinct aspect of output quality. 
By measuring both per-metric performance and overall success rates, it enables fine-grained analysis of \orion{}'s behavior while reducing reliance on repeated manual assessment.

Evaluation relies on human-annotated ground truth, with different strategies depending on output type. 
For numerical outputs, we compare system results directly to ground truth values, applying a tolerance of 0.5 to account for floating-point precision and off-by-one variation. 
For textual outputs, we assess semantic equivalence against ground truth using LLM judges, which compare candidate outputs to reference statements and determine correctness.

We note that LLM judges are only reliable when reference ground truth is available; without such standards, judgments can be inconsistent. 
Finally, because interface identification and harness generation present distinct challenges, we define custom evaluation metrics for each task, as detailed in the following subsections.

\subsection*{Interface Identification Metrics}
The evaluation metrics for interface identification align with those used by \orion{}'s analysis agent. 
They fall into three categories:

\begin{itemize}[leftmargin=*]
  \item \textit{Numerical metrics:} Lines of Code, Cyclomatic Complexity, Internal Calls, and Callgraph Size. 
  These are compared directly against ground truth values, with minor tolerance applied for floating-point precision.

  \item \textit{Qualitative metrics:} Sink Functions, Dangerous Expressions, and Parsing Functions. 
  These produce natural-language outputs that are evaluated against annotated ground truth using LLM judges for semantic equivalence.

  \item \textit{Auxiliary classification:} Identification of functions with no parameters, reported as a binary accuracy metric.
\end{itemize}

\subsection*{Fuzz Harness Quality Metrics}
The harness generation benchmark evaluates \orion{}'s outputs against eight criteria that reflect the requirements for functional and effective fuzz harnesses (Section~\ref{sec:fuzzing_101}). 
All metrics are assessed using LLM judges against human-annotated ground truth.

\begin{itemize}[leftmargin=*]
  \item \textit{Verification}: Checks whether harnesses pass self-reflection and compiler validation. Only compilable and semantically valid harnesses proceed to fuzzing.

  \item \textit{Input Channels}: Ensures harnesses accept input through all relevant channels (e.g., parameters, globals, files).

  \item \textit{Input Structure}: Evaluates correct transformation of fuzzer input into the target interface format, including appropriate type usage and population.

  \item \textit{Dependencies}: Confirms correct program setup, including initialization routines and dependency handling. Missing dependencies are counted as failures, as they prevent successful execution.

  \item \textit{Teardown}: Validates proper resource cleanup and program state reset after each iteration to avoid contamination and performance degradation.

  \item \textit{Unnecessary Functions}: Flags operations outside the target interface, setup, teardown, or dependencies that reduce fuzzing throughput.

  \item \textit{Proper Use of Fuzzed Data}: Checks that input is derived correctly from fuzzer data, with required constants or checksums incorporated appropriately.

  \item \textit{Target Function Invocation}: Ensures the harness ultimately calls the intended target interface, even after iterative refinement and error correction.
\end{itemize}




\subsection{RQ1: Interface Identification Metrics}
\label{sec:rq1}

\begin{table}[t]
    \centering
    \small
    \begin{tabular}{|l|r|r|r||r|}
        \toprule
        \textbf{Metric} & \multicolumn{1}{c|}{\textbf{clib}} & \multicolumn{1}{c|}{\textbf{H3}} & \multicolumn{1}{c||}{\textbf{GPIO}} & \multicolumn{1}{c|}{\textbf{Average}} \\
        \midrule
        \rowcolor{gray!20} Lines of Code & 100.00\% & 100.00\% & 100.00\% & 100.00\% \\
        Cyclomatic Complexity & 100.00\% & 100.00\% & 100.00\% & 100.00\% \\
        \rowcolor{gray!20} Internal Calls & 100.00\% & 100.00\% & 100.00\% & 100.00\% \\
         Callgraph Size & 100.00\% & 100.00\% & 100.00\% & 100.00\% \\
        \rowcolor{gray!20} Sink Functions & 86.21\% & 96.97\% & 93.33\% & 94.30\% \\
        Dangerous Expressions & 89.66\% & 75.76\% & 66.67\% & 76.58\% \\
        \rowcolor{gray!20} Parsing Functions & - & 100.00\% & 100.00\% & 100.00\% \\
        No Arguments & 100.00\% & 100.00\% & 60.00\% & 85.71\% \\
        \midrule
        \rowcolor{gray!20} Overall & 95.42\% & 96.11\% & 93.49\% & 95.48\% \\
        \bottomrule
    \end{tabular}
    \caption{Benchmark Results For Interface Identification Metrics}
    \label{tab:benchmark_results_rq1}
    \vspace{-1em}
\end{table}

Table~\ref{tab:benchmark_results_rq1} reports \orion's performance on the interface identification task. 
Across all metrics and projects, \orion achieves an average success rate of 95.5\%. 
The system attains perfect scores on the four numerical metrics (\texttt{Lines of Code}, \texttt{Cyclomatic Complexity}, \texttt{Internal Calls}, and \texttt{Callgraph Size}), indicating reliable integration of tool-based measures with LLM reasoning. 
High success rates on the remaining metrics reflect the benefits of reliability techniques such as self-consistency, self-reflection, reasoning-focused prompts, and precise code retrieval.

We identify three primary sources of error in the more challenging metrics (\texttt{Sink Functions} and \texttt{Dangerous Expressions}):  
(i) \textbf{Insufficient context specificity}, particularly in H3 where heavy macro usage makes code harder to resolve. Providing too much context risks the needle-in-the-haystack effect (Section~\ref{sec:llms}), while too little context omits critical information.  
(ii) \textbf{Counting and dereference errors}, including miscounts of sink functions and missed pointer dereferences. These stem from known LLM weaknesses with arithmetic and limited contextual resolution.  
(iii) \textbf{Self-reflection pitfalls}, where incorrect initial analyses occasionally persisted because the review stage accepted, rather than corrected, flawed reasoning. 

Although carefully crafted prompts and feedback loops mitigate most failures, these results highlight areas where LLM-based analysis remains vulnerable, underscoring the importance of validation and tool support.


\subsection{RQ2: Fuzz Harness Quality}
\label{sec:rq2}

\begin{table}[t]
    \centering
    \small
    \begin{tabular}{|l|r|r|r||r|}
        \toprule
        \textbf{Metric} & \multicolumn{1}{c|}{\textbf{clib}} & \multicolumn{1}{c|}{\textbf{H3}} & \multicolumn{1}{c||}{\textbf{GPIO}} & \multicolumn{1}{c|}{\textbf{Average}} \\
        \midrule
        \rowcolor{gray!20} Verification & 100.00\% & 100.00\% & 42.42\% & 88.20\% \\
        Input Channels & 93.10\% & 97.98\% & 100.00\% & 97.52\% \\
        \rowcolor{gray!20} Input Structure & 82.76\% & 90.91\% & 60.61\% & 83.22\% \\
        Dependencies & 48.28\% & 86.87\% & 27.27\% & 67.70\% \\
        \rowcolor{gray!20} Teardown & 89.66\% & 100.00\% & 33.33\% & 84.47\% \\
        Unnecessary Functions & 96.55\% & 97.98\% & 90.91\% & 96.27\% \\
        \rowcolor{gray!20} Proper Use of Fuzzed Data & 82.76\% & 94.95\% & 100.00\% & 93.79\% \\
        Target Function Invocation & 100.00\% & 100.00\% & 100.00\% & 100.00\% \\
        \midrule
        \rowcolor{gray!20} Overall & 86.64\% & 96.09\% & 69.32\% & 88.90\% \\
        \bottomrule
    \end{tabular}
    \caption{Benchmark Results For Fuzz Harness Quality Metrics}
    \label{tab:benchmark_results_rq2}
    \vspace{-1em}
\end{table}

\orion achieves an average success rate of 88.9\% across harness quality metrics, making this task more challenging than interface identification. 
The lowest scores appear in \texttt{Dependencies} (67.7\%) and \texttt{Teardown} (84.5\%), reflecting the inherent difficulty of reasoning about initialization and cleanup requirements from source code - a task that is error-prone even for human experts. 
In contrast, \orion performs strongly on \texttt{Verification}, \texttt{Input Channels}, and \texttt{Target Function Invocation}, showing that iterative refinement and compiler feedback reliably enforce basic correctness.

Error analysis reveals three main challenges. 
First, dependency handling remains brittle: missing initialization or teardown steps frequently led to invalid harnesses. 
Second, harnesses often failed compilation on first attempt, making iterative refinement through compiler feedback and self-reflection essential. 
Finally, the GPIO driver results illustrate the difficulty of scaling to large proprietary codebases with unfamiliar build systems. 
Here, \orion frequently produced incorrect header assumptions and misinterpreted compiler diagnostics, reducing verification accuracy.

Overall, these results show that while \orion can reliably generate functional harnesses in many cases, dependency analysis and complex build environments remain key limitations that warrant further research.

\subsection{RQ3: Time Savings}
\label{sec:rq3}

\begin{table}[t]
    \centering
    \small
    \begin{tabular}{|l|r|r|r|}
        \toprule
        \textbf{Task} & \textbf{Human Effort} & \textbf{Orion Effort} & \textbf{Speedup} \\
        \midrule
        \rowcolor{gray!20} Interface Identification & $\sim$1 week & 62 min & 92$\times$ \\
        Harness Generation & $\sim$2 weeks & 49 min & 204$\times$ \\
        \rowcolor{gray!20} Patching & 1 hr & 1 min 20 sec & 46$\times$ \\
        \bottomrule
    \end{tabular}
    \caption{Time savings achieved by Orion compared to human effort.}
    \label{tab:rq3_timesavings}
    \vspace{-1em}
\end{table}

To quantify \orion's practical benefit, we measured the time required for interface identification, harness creation, and vulnerability patching, using NVIDIA internal data for tasks comparable to the clib zero-days. 

As shown in Table~\ref{tab:rq3_timesavings}, Orion reduces workflows that typically take days or weeks of analyst effort to less than two hours of automated execution. 
Harness generation yields the largest speedup, reflecting the high manual burden of crafting harnesses through detailed program analysis and iterative refinement. 
Interface identification achieves a smaller (yet still substantial) gain, as it must process significantly more candidate interfaces. 
Here, the use of self-consistency increases runtime by requiring multiple parallel LLM queries, introducing a deliberate speed–accuracy tradeoff. 
Finally, patching also provides large acceleration, but still requires human verification of candidate patches, limiting potential speedup relative to earlier workflow stages.


\subsection{RQ4: Zero-day Vulnerabilities}
\label{sec:rq4}
The ultimate measure of \orion's effectiveness is whether its outputs can lead to the discovery of new, previously unknown vulnerabilities. 
To evaluate this, we applied \orion to the top 12 candidate functions identified in the H3 and clib codebases, executing each harness for up to 24 hours or until a crash was detected.

This evaluation yielded two previously unknown vulnerabilities in \texttt{clib}: a controlled stack buffer overflow and a null pointer dereference. 
Both issues were discovered automatically through \orion's generated harnesses and seeds, demonstrating that the system can not only reproduce known bug-finding workflows but also surface new security-critical defects.

At the time of writing, both vulnerabilities remain under coordinated disclosure. 
Full technical details are therefore withheld, but will be released once responsible disclosure has concluded.

\section{Discussion}


The code knowledge base serves as \orion{}'s crucial foundation, enabling precise context retrieval for LLM tasks such as symbol definitions and function callers.
While these tasks resemble those of Language Server Protocols (LSPs)~\cite{lsp} used by modern IDEs, existing LSPs require specialized wrappers to support AI workflow requirements.
Creating LSP wrappers targeted at AI systems could provide valuable community building blocks and could unlock the full potential of LLMs for code analysis.

Build system interactions present significant challenges due to diverse implementations across projects, even within identical build systems.
This variety complicates harness integration into the codebase and tasks requiring build configuration knowledge, such as header resolution for symbols.
\orion currently addresses this incompletely, as build system understanding represents an orthogonal problem to fuzzing automation.
Solutions involve either requiring detailed user inputs (reducing usability) or developing specialized build system analysis components (increasing complexity but lowering barriers), where LLM agents show promise.

\orion{}'s modular design intentionally decouples components to enable isolated usage and flexible adaptation for special cases.
This modularity facilitates experimentation with different approaches across individual workflow steps and multi-step combinations, with Section~\ref{sec:related_work} identifying promising research directions.

\section{Related Work}
\label{sec:related_work}
In this section we describe prior efforts in the area of scaling fuzzing as well as the use of LLMs for security tasks.

Prior work relevant to Orion falls into three areas: (i) applying LLMs to fuzzing, (ii) scaling fuzzing workflows without generative AI, and (iii) broader AI applications in software security.

\textbf{LLMs for Fuzzing.} 
Recent research has applied LLMs to different components of the fuzzing pipeline. 
\textit{End-to-end systems:} The closest effort is DARPA’s AIxCC challenge~\cite{aixcc}, where finalist teams built systems for vulnerability discovery and patching across diverse codebases. Unlike \orion, these systems focused on general vulnerability detection rather than workflow automation, and relied on organizer-provided harnesses.  
\textit{Single-step automation:} Harness generation was first demonstrated by \citeauthor{googleHarnessGeneration}~\cite{googleHarnessGeneration} through OSS-Fuzz integration, followed by iterative refinement~\cite{promptFuzzingFuzzDriver, elfuzz} and context-aware prompting~\cite{ckgFuzzer, deepSURF, llmFuzzDriverGeneration, kernelgpt}. Seed generation has been explored via refinement~\cite{seedmind}, targeted context~\cite{psgSeedGen}, and constraint solving~\cite{hgfuzzer}. For crash triage and patching, CodeRoverS~\cite{coderovers} showed that minimally customized agents can generate fixes for fuzzing-induced crashes. All of these focus on isolated workflow steps, whereas \orion automates the complete pipeline from target identification through patching.  
\textit{Closed-source fuzzing:} LibLMFuzz~\cite{liblmfuzz} applies agentic workflows with disassemblers, compilers, and fuzzers to fuzz binary-only targets.  
\textit{LLM-enhanced fuzzing engines:} A separate line of work uses LLMs to augment fuzzing itself, including input generation and mutation~\cite{chatafl, fuel, llamafuzz}, constraint solving~\cite{hyllfuzz, hlpfuzz}, and engine replacement (e.g., Fuzz4All~\cite{fuzz4all}). These efforts complement \orion but focus on fuzzing mechanics rather than workflow automation.

\textbf{Scaling Fuzzing without Generative AI.} 
Before LLMs, researchers explored traditional automation for individual workflow steps. 
Target identification relied on static metrics~\cite{sokFuzzTargets}. 
Harness generation used unit tests~\cite{utopia}, static analysis, or dynamic tracing~\cite{fuzzgen, wildsync, winnie}. 
Seed generation applied structural learning, grammars, or ML-based feature extraction~\cite{skyfire, smartseed, mlseed}. 
Crash analysis employed dynamic tracing for invariant learning~\cite{aurora} and grouping methods for root cause identification~\cite{igor}. 
Automated patching leveraged genetic programming~\cite{genprog} or symbolic execution with program synthesis~\cite{semfix, angelix, pmbugassist}. 
While valuable, these methods addressed only single workflow phases, often without semantic understanding, and produced outputs that were difficult to maintain in practice.

\textbf{AI for Security.} 
Beyond fuzzing, generative AI has enabled a range of security applications. 
Big Sleep~\cite{bigSleep} (formerly Project Naptime~\cite{projectNaptime}) performs variant analysis on large codebases. 
XBOW~\cite{xbow} applies agentic AI to web application penetration testing. 
Vul-RAG~\cite{vulrag} leverages vulnerability history for LLM-powered vulnerability detection. 
Other work augments static analysis, such as reducing false positives~\cite{llift, chatgptstatic} and automatically generating checkers to improve scalability~\cite{knighter}. 
These advances highlight the growing role of AI in software security and motivate \orion's focus on end-to-end fuzzing workflows.

\section{Conclusion}

We introduced \orion, an end-to-end framework for fuzzing workflow automation that enables scalable campaigns on production-size codebases. 
\orion combines LLM-based agents with deterministic tools to automate the complete pipeline from target identification through patch generation. 
The hybrid design exploits LLM strengths in semantic reasoning and code understanding while relying on traditional tooling for verification, iterative refinement, and precise context retrieval.

Evaluation results demonstrate that \orion delivers both effectiveness and efficiency: it achieved 95.5\% accuracy in interface identification, 88.9\% in harness generation, reduced manual effort by up to 204$\times$, and uncovered two previously unknown vulnerabilities. 
These findings highlight the feasibility of integrating LLM agents with conventional techniques for security-critical software testing.

Important challenges remain. 
Improved dependency analysis, support for complex build environments, and more rigorous patch validation represent promising directions for future work. 
Addressing these challenges can further increase the reliability and adoption of automated fuzzing workflows in large-scale software development.

\begin{acks}
We thank Tom McReynolds for supporting this project.
\end{acks}

\bibliographystyle{ACM-Reference-Format}
\bibliography{orion}

\end{document}